\documentclass[12pt]{article}
\usepackage[margin=1in]{geometry}
\usepackage{graphicx}
\usepackage{float}
\usepackage{microtype}
\usepackage{lmodern}
\usepackage[T1]{fontenc}
\usepackage{hyperref}
\usepackage{etoolbox}
\usepackage{mathtools}
\usepackage{cleveref}
\usepackage{booktabs}
\usepackage{multirow}
\usepackage{minted}

\usepackage{siunitx}
\usepackage[authordate,bibencoding=auto,strict,backend=biber,natbib]{biblatex-chicago}
\begin{document}

\title{Visualization Supporting Talent Portrait Generation: an Empirical Study}
\author{Yuqing Fan$^{1,2}$, Shenghui Cheng$^{2}$\\
$^1$ Chongqing Open University, Chongqing, China \\
$^2$ Jinan University, Guangzhou, China
}
\date{}

\maketitle

\begin{abstract}

In today’s era of scientific and technological advancements, the importance of talent resources is increasingly highlighted. This article will attempt to summarize the academic trajectories and successes of numerous researchers from both past and present, aiming to reproduce the correlation between their personal development and academic output. This article propose a framework of academic data collection, then use the framework to visualise numerous scientists' research accomplishments, collaborative partners, and research inheritance. The article additionally posited a range of scholarly indices designed to quantitatively assess the impact of scientists.

\end{abstract}

\section{Introduction}

The development of science is not only about the constant updating of scientific theories and technological means, but also the process of inheritance and development of scientific knowledge, scientific traditions, and scientific culture among generations of scientists. Scientists at the forefront of the world often have distinct teacher-student relationships. The history of academia is derived from academia. Therefore, the nature of academia determines the nature of its history (\cite{zlw2004}).

With the emergence of talent shortages, brain drain, and structural contradictions in employment, the use of visual methods for human resource management has become increasingly popular in describing talent development. The "14th Five Year Plan" Big data Industry Development Plan proposes to build a prosperous and orderly industrial ecology and improve the level of Big data public services such as talent training. Many local governments, enterprises, and institutions also use visual and quantitative features to search for talents that meet their needs. In this article, we suggest combining visualization methods with researchers' career trajectories to analyze their careers from three perspectives: (1) research projects led by the researchers themselves, (2) collaborative projects in which the researchers participate, and (3) teacher-student relationships between researchers. In addition, this article will quantitatively study the output of researchers through longitudinal analysis and use it to predict the future development of researchers.

\section{Research status}

Currently, very few paper had considered visualization techniques to evaluate scientists themselves and quantitatively analyze their achievements at the same time. However, individual topics have been studied. For example,  \cite{sokolov2020} evaluate the extent of researchers' involvement in various fields through an evaluation graph with eight dimensions, including social sciences, natural sciences, and engineering. Then they use the t-sne visualization method to evaluate the similarity between different researchers' studies. \cite{maceachren2004} integrate scientists in the field of geology with scientific discoveries, terms, and concepts, helping future geologists learn and research through spatial data and process diagrams. \cite{qxy2017} use the Gartner model, business process diagrams, and T-type evaluation diagrams to evaluate the capabilities and research status of foreign data scientists.

In terms of academic inheritance research, various departments and faculties of universities around the world have commemorative webpages presenting inheritance. For example, the University of Manchester commemorates the school's history in the school history museum, including the research process and achievements of 25 Nobel Prize winners since the Victorian era, as well as the research achievements of famous researchers in the school's history. European higher education organizations also document a large number of great achievements and inventions, as well as famous researchers and their lineage, in famous universities in Europe since the establishment of the University of Bologna in 1088. There are also studies on academic inheritance topics in China. \cite{lzf2012} discusses the academic inheritance of Chinese scientists from the perspectives of learning Western modern science and rediscovering the traditional academic values of China.

Quantitative methods have become a popular method in recent years to evaluate the performance of candidates. At present, the most popular method in China is the study of competency models. It is generally believed that competence is a personal condition and behavioral characteristic that directly affects work performance, referring to the deep-seated characteristics of individuals who can distinguish outstanding achievements from ordinary people in a certain job \cite{ly2006}. \cite{xf2012} proposed a human resource performance management system based on personal competence. Every employee must possess corresponding abilities to achieve their job performance goals. In addition, \cite{lsc2008}, \cite{clw2006}, and \cite{hyx2008} had all discussed the practice and application of competency models from different perspectives. In addition to the competency model, \cite{yzl2020} classified the current domestic intellectuals into six categories: basic research talents, applied research talents, technology development talents, achievement transformation talents, technology management service talents, and experimental technology talents. They also created a user tag based on all published journal, conference, and patent results, to evaluate a talent from different perspectives. Of course, there are also many domestic studies scoring talents in the form of Indexation. For example, Qiu Junping and Miao Wenting used the Hirsch index to evaluate researchers in the field of library and information science \cite{qjp2007}. In addition to the h index, the Ht index proposed by \cite{xx2014}, the Prathap index proposed by \cite{wzj2012}, and the RAC index proposed by \cite{flm2019} all try to evaluate talents in a quantitative way.

\section{Framework of data visualization}

As the size of companies expand, their human resource departments usually demand more complex management methods and tools. The commonly used quantitative analysis methods include comparative analysis, attribute analysis, and graphic analysis. Comparative analysis and comparison of data to demonstrate differences between different data groups; Attribute analysis focuses on studying the changes between things, tending towards traditional quantitative analysis; Graphic analysis utilizes graphics as a means to provide readers with a better reading experience (\cite{rkl2019}).

In the past, human resource management in universities mainly focused on managing the current situation of human resources, lacking foresight. Currently, due to limited information, human resource services have become increasingly passive. \cite{zln2016} believe that there are five main problems in current human resource management in universities: the inability to achieve precise matching between positions and in-service personnel, as well as potential recruitment talents (\cite{ly2015}); The quantification level of performance evaluation indicators is not high (\cite{ll2011}); The information is limited and cannot be predicted; The phenomenon of brain drain is widespread (\cite{qd2014}); Human resource management is too rigid to meet individual needs (\cite{gys2015}).

Therefore, this article proposes a framework map for analyzing data from researchers. For any government agency or enterprise or institution that wants to recruit, visual charts can be created, compared, and evaluated through the above methods.

\subsection{Data collection and classification}

There are a large number of platforms at home and abroad that provide scientific research data for researchers, especially journals and related data. Many researchers themselves create personal interfaces or provide relevant personal information on school platforms. Public platforms such as CNKI, Google Academic, Research-Gate, and Web of Science also provide specific data on the authors, citations, and downloads of journal and conference papers.

\begin{figure}[H]
    \centering
    \includegraphics[scale=0.5]{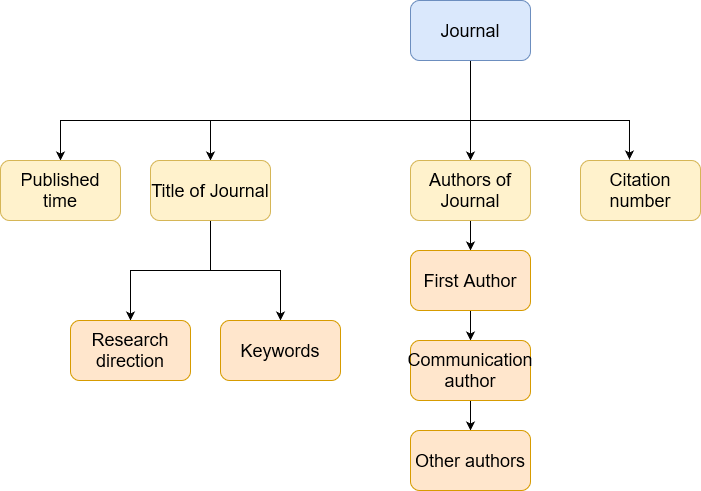} 
    \caption{Framework of data collection}
    \label{123en}
\end{figure}

In the process of data collection, a common situation is that the publication time of the paper on the personal interface does not match the data platform. This is due to the fact that the publication process of the paper spans years, resulting in different information between the two. In this case, the publication time of the personal interface should prevail. The detailed framework for data collection is shown in Figure \ref{123en}.

Various methodologies exist for gathering data on academic lineage. For researchers in the fields of mathematics or computer science, direct inquiries can be made via platforms such as Mathematical Genealogy Engineering. However, it is important to note that these repositories exclusively provide information pertaining to individuals within these specific disciplines. For researchers operating outside of these domains, alternative data collection methods become requisite. These may involve accessing the researcher's personal interface, consulting the institution's paper database, and utilizing departmental interfaces.

Following the data acquisition phase, a systematic classification process ensues, wherein all publications associated with the research subject are categorized into two primary groups: those constituting research projects led by the researcher themselves, and those representing research projects in which the researcher participated. The present study posits that, in the majority of instances, when the first author of an article is the research subject or when the corresponding author is the research subject, the research can be ascribed to a project led by the researcher. Conversely, the inverse scenario holds true.

\subsection{Framework for analysis of Journals}

The purpose of establishing a literature analysis framework in this article is to reflect the changes in the quality of papers while reflecting the number of papers published each year. Therefore, this article adopts a columnar line overlay chart to simultaneously include two sets of data.

The innovation of this article lies in the addition of the label of more influential papers, taking into account data dimensionality reduction, to demonstrate the proportion of outstanding papers by researchers in the total number of papers per year (\cite{wang2019two}). Due to the different number of factors in journals of different disciplines, the proportion of high factor papers varies among disciplines. Taking statistics as an example, journals with factors ranging from 4 to 5 are usually considered influential. Therefore, excellent papers can include papers published in journals with high impact factors, as well as papers with high citations. However, overall, the proportion of excellent papers should be around 10-20\%, and should not exceed 20\%. In the future, the improvement direction of the literature analysis framework is to consider more elements of high-dimensional analysis (\cite{zang2022evnet}).

In the end, this article proposed a longitudinal model to estimate the overall effect of how scientists evolved through time. The model will consider the cumulative numbers of journals which are considered more important as an indicator through time, and thus the model will be able to test the difference among scientists.

\subsection{Framework for analysis of co-authors}

The collaborators of researchers include all research projects that involve researchers, and this article analyzes these projects in two different ways. Firstly, this article analyzed the working institutions of all participants in these projects and conducted spatial data analysis on these institutions. Subsequently, the article analyzed the cooperation between participants in each project, except for the research subjects, and found the frequency of academic cooperation from two aspects. In the process of visualization, we should focus on the ethnic level of academic cooperation, and reflect the number of cooperation in the form of thick and thin lines. In the future, collaborators can also conduct clustering analysis of data texts (\cite{7194836}), or multi factor analysis using different colors for labeling (\cite{Huang2023HighdimensionalCO}).

\subsection{Framework for analysis of academic inheritance}

Academic inheritance analysis also includes two different visualization methods. The first part shows the Doctoral advisor of the research goal and the upward inheritance. In the process of visualization, it is necessary to collect the birth and death dates of each researcher on the inheritance tree, so as to build a complete Gantt chart. The second section displays the graduation destinations of all doctoral students under the guidance of the researchers, and displays them through spatial data. It should be noted that the visualization process should include students heading to work in the industry, and each student's position should be processed according to the latest position.

On the other hand, based on the students' influence index and their distance to the home university, this article proposed a cumulative index of all students of a researcher. The index can be used to consider the teaching ability of scientists, rather than simply compare number of students.

\subsection{Dealing with missing data}

Missing data is a very common phenomenon in data processing. Generally speaking, when there is a missing data, the most common approach is the data filling method, which converts a dataset with missing data into a complete dataset by filling in the mean, mode, or classification values such as 1 and -1 (\cite{chen2023general}). Within the framework of this article, common missing data values include: impact factors of journals that cannot be queried, graduates that cannot be queried, and the biographies of ancient scientists. Due to the importance of these values, the use of data filling method is not reasonable. Generally speaking, as long as the total missing rate of each data item is not higher than 5\%, only complete data can be analyzed (\cite{jakobsen2017}).

\section{Data Visualization}

This section will introduce our visualization works to describe the efforts of scientists, we will estimate a faculty's work with 3 different perspectives: 1) Journals published in recent years, 2) Collaborators in the world and 3) Academic Inheritance.

\subsection{Analysis of Journals}

\begin{figure}[H]
    \centering
    \includegraphics[scale=0.7]{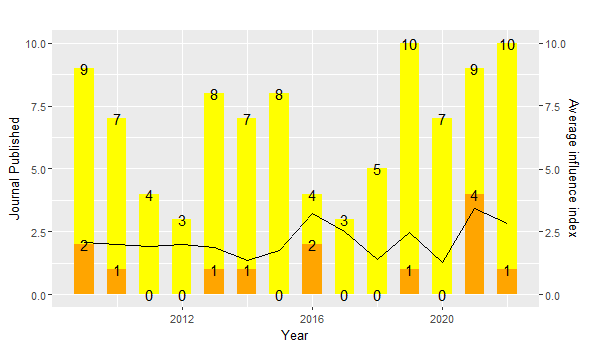} 
    \caption{Journal analysis of the selected Professor}
    \label{cook1}
\end{figure}

As shown in Figure \ref{cook1}, it displays all the papers published by someone from the Department of Actuarial Science and Statistics at the chosen university. The figure presents several significant papers published by the person, which are either published in high-impact factor journals or have gained a large amount of application.

From the graph, it is not difficult to see that the number of articles published by the person has been around 5 to 6 per year since 2009. Among them, the number of publications clearly shows differences with troughs and peaks. The years 2011-2012 and 2016-2017 had a lower number of articles published. However, the average impact factor of the published articles has shown a gradual increase. It has remained stable at around 2.5 in the early years and has been increasing in recent years, reaching a peak of 4 in one year. This is related to the fact that journals in the field of statistics generally have lower impact factors.

\begin{figure}[H]
    \centering
    \includegraphics[scale=0.7]{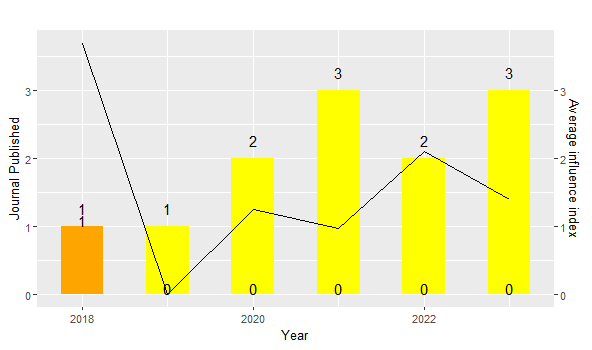} 
    \caption{Journal analysis of selected PhD student}
    \label{lqd}
\end{figure}

As a comparison, we selected one doctoral graduate of the selected person for comparison. Figure \ref{lqd} shows the publication status of the doctoral graduate papers in the past six years. The data in recent years has also shown an upward trend.

With a framework of analysis of journal presented, it is possible to conduct a longitudinal analysis across different scientists. As a result, a linear mixed effect model maybe applied. Notice that the ratio of important articles seems to remain at a value suppose the total articles reached a level. 

\begin{equation}
\label{eq1}
Y_{ij}=X_{ij}^{'}\beta+X_{ij}^{'}b_{i}+\varepsilon_{ij}
\end{equation}

Within the framework of the linear mixed-effects model delineated herein, \(X_{ij}\) denotes the cumulative count of journals regarded as more significant over successive years. The error term \(\varepsilon_{ij}\)  is posited to follow a normal distribution and is presumed to be identically distributed, given its nature as a manifestation of individual measurement variability. With estimation of \(\beta\) and \(b_{i}\), it becomes feasible to ascertain the values corresponding to population-level effects and individual-level heterogeneity.

Due to the parametric nature of LME model with normality assumed. The maximum likelihood can be directly computed with asymptotic equation:

\begin{equation}
\label{eqasy}
\hat{\beta} \sim N(\beta,[\sum_{i=1}^{n}X_{i}^{'}V_{i}^{-1}(\theta)X_i]^{-1})
\end{equation}

The variance in the equation is given by \(V_{i}(\theta)=Z_{i}DZ_{i}^{'}+\sigma^2I\). D is the variance of covariate b. After get the value of beta, a Likelihood ratio test can be done since the method is based on likelihood functions.

\subsection{Analysis of co-authors}

\begin{figure}[H]
    \centering
    \includegraphics[scale=0.7]{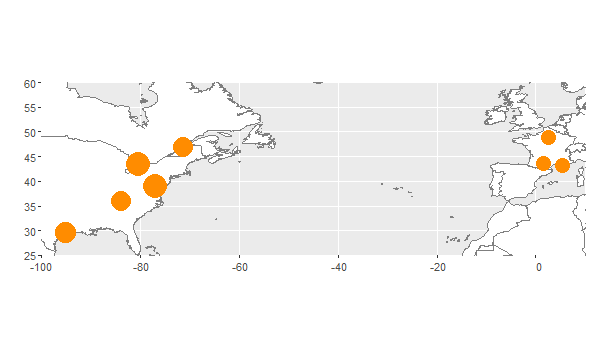} 
    \caption{Spatial data of co-authors of the selected professor}
    \label{rcco}
\end{figure}

Illustrated in Figure \ref{rcco}, the visual representation delineates the scholarly collaborations of the selected professor. The selected professor's principal collaborators emanate predominantly from educational institutions in both Canada and the United States, while a limited subset also originates from schools and institutions in France.

\begin{figure}[H]
    \centering
    \includegraphics[scale=0.9]{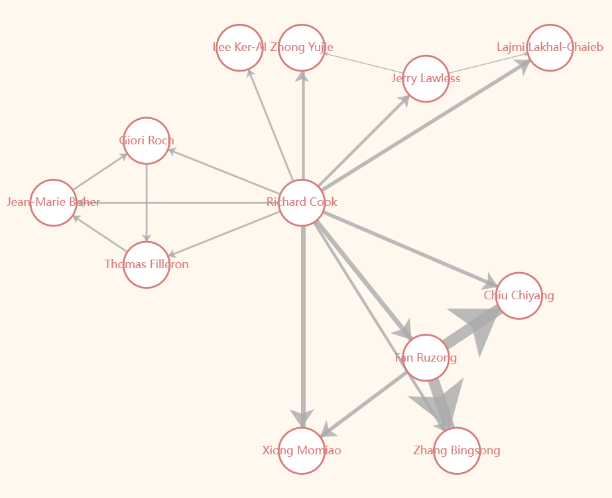} 
    \caption{Analysis of co-authors }
    \label{rcco2}
\end{figure}

The subsequent Figure \ref{rcco2} delineates the precise individuals who have collaborated. The findings conspicuously unveil the compositional distribution of these collaborators, with a predominant contingent comprising individuals affiliated with the University of Waterloo, primarily encompassing students and colleagues. The left segment of the graph depicts an additional academic cohort from Aix-Marseille University in France, while the upper-right quadrant illuminates discernible connections to a cohort of Chinese academics operating within United States-based universities.

In order to assess the cumulative impact of academic collaborations, it is conceivable to introduce a novel index designed to encapsulate the average influence exerted by journals resulting from collaborative endeavors at either the national or international level. The formula for this index is expressed as follows:

\begin{equation}
\label{eq2}
I(\text{Coop})=\sum_{i=1}^{n} \frac{c_i t_i}{n}
\end{equation}

The recently proposed index computes the mean influence attributed to a journal. In the formula, the variable \(c_i\)  denotes the citation count of the article, and \(t_i\)  represents the geographical distance between the respective home universities of the collaborating authors. In broad terms, assuming an author is capable of engaging in collaborative efforts at a continental or transcontinental scale, the associated journal is deemed to exert a heightened impact, thereby signifying the author's proficiency in establishing expansive scholarly networks.

\subsection{Analysis of academic inheritance}

Presently, numerous scholarly initiatives are dedicated to the historical reconstruction of academic lineage and the delineation of intellectual connections between scholars. Notably, the "Mathematics Genealogy Project" has undertaken the systematic organization of genealogical information pertaining primarily to mathematicians, spanning epochs from ancient times to the contemporary era.

\begin{figure}[H]
    \centering
    \includegraphics[scale=0.7]{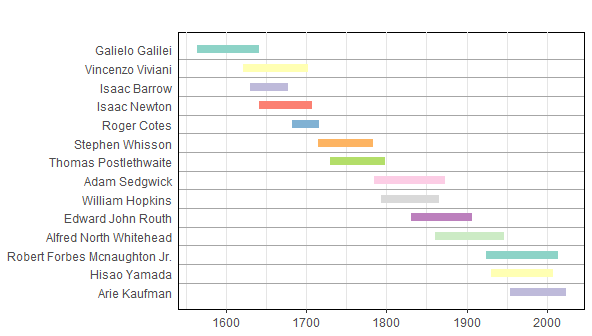} 
    \caption{Analysis of academic antecedents of Ari Kaufman}
    \label{ari}
\end{figure}

The objective of this article is to visually depict the interconnections among scholars based predominantly on data procured from the Mathematics Genealogy Project. Illustrated in Figure \ref{ari}, an analysis of academic inheritance is conducted, employing Professor Ari Kauffman from Stony Brook University as an illustrative case. It is pertinent to highlight that the compiled list encompasses esteemed figures such as Newton and Galileo. This recurring pattern, observed in the genealogical project, reveals a common phenomenon where, through meticulous tracing, one's academic lineage can be discerned to extend back to eminent names such as Leibniz and Newton.

\begin{figure}[H]
    \centering
    \includegraphics[scale=0.7]{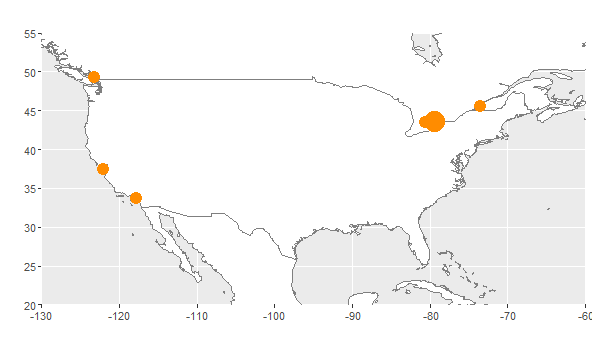} 
    \caption{PHD students’ jobs of David Andrews Francis}
    \label{daf}
\end{figure}

Beyond the examination of academic lineage, this article introduces a visual methodology for scrutinizing the post-doctoral trajectories of students. Illustrated in Figure \ref{daf}, the diagram delineates the destinations chosen by 17 doctoral graduates who were under the mentor-ship of someone within the Department of Statistics at the University of Toronto. The visual representation illustrates the preferred career paths pursued by these graduates, with a predominant inclination towards academic positions in institutions located in the United States and Canada. Additionally, a cohort of graduates has chosen to transition into the industrial sector.

Ultimately, a comprehensive academic lineage influence index may be formulated by aggregating the individual accomplishments of students. With each student's H-index called \(h_j\) and relative distant \(t_j\), the total academic achievement equation is given by:

\begin{equation}
\label{eq3}
I(\text{Inherit})=\sum_{i=1}^{n} h_i t_i
\end{equation}

\section{Conclusion}

This paper undertakes a visual scrutiny of extant researchers' research accomplishments, yielding the following deductions: foremost, the visual analysis of researchers emerges as a pivotal undertaking. By transmuting copious data and information into graphical representations, it facilitates a more intuitive comprehension of various dimensions of researchers, encompassing academic attainments, current scholarly standing, international collaboration dynamics, and the distribution of students.

Concerning the indices propounded in this exposition, avenues for refinement are discernible. Notably, the distance covariate in both the inheritance and cooperation indices fails to distinctly reflect the substantial international impact engendered by journals. Consequently, future iterations may benefit from the incorporation of an additional covariate signifying university ranking to enhance the comprehensiveness of the indices.

As emerging scholars undergo maturation and contribute progressively to academic discourse, it behooves major educational institutions to engage in prospective prognostications. This strategic foresight aids in cultivating a preliminary understanding of the trajectory of young researchers, enabling more informed strategic planning for both institutional and individual scholarly development. Looking ahead, with the advent and progression of technologies such as the Metaverse and Web 3.0, the visual analysis of scientific researchers is poised to evolve, promising a more nuanced and accurate depiction of their research proficiency and achievements (\cite{cheng2022roadmap,fan2023current,chengbook2023,huang2023roadmap}).

\newpage 
\printbibliography[title=Bibliography]

\end{document}